\documentclass[3p,times,twocolumn]{elsarticle}

\usepackage{ecrc}
\usepackage{amsmath}
\usepackage{amssymb}
\usepackage{amsfonts}

\volume{00}

\firstpage{1}

\journalname{Nuclear Physics B Proceedings Supplement}

\runauth{}

\jid{nuphbp}

\jnltitlelogo{Nuclear Physics B Proceedings Supplement}

\usepackage{amssymb}

\usepackage[figuresright]{rotating}

\begin{document}

\begin{frontmatter}



\dochead{}

\title{\bf Low mass right-handed gauge bosons from minimal grand unified theories}


\author{\bf Biswonath Sahoo and M. K. Parida}

\address{Centre of Excellence in Theoretical and Mathematical Sciences, \\
Siksha 'O' Anusandhan University, Khandagiri Square, Bhubaneswar 751030, \\
Odisha,  India}
\ead{minaparida@soauniversity.ac.in}
\begin{abstract}
    Prediction of low-mass $W_R$ and $Z_R$ gauge bosons in popular grand unified theories has
been the subject of considerable attention over the last three
decades. In this work we show that when gravity induced
 corrections due to dim.5 operator are included the minimal symmetry breaking chain of $SO(10)$ and $E_6$ GUTs can yield $W_R^{\pm}$
and  $Z_R$ bosons with masses in the range $(3-10)$ TeV which are
accessible to experimental tests at the Large Hordan Collider.  
The RH neutrinos turn out to be heavy pseudo-Dirac fermions.
The model can fit all fermion masses and manifest in rich structure of lepton flavor violation while proton life time 
is predicted to be much longer than the accessible limit of
Super-Kamiokande or planned Hyper-Kamiokande collaborations.
\end{abstract}


\end{frontmatter}


\section*{}
 Left-right symmetric gauge theory \cite{pati,rnm} originally suggested to explain parity violation as monopoly of weak interaction has also found 
 wide applications in many areas of particle physics beyond the standard model including neutrino masses and mixings, lepton number and lepton flavor 
 violations,CP-violation , $K-\bar{K}$ and $B-\bar{B}$ mixings and baryogenesis through leptogenesis. This theory is expected to make substantial new  
 impact on weak interaction phenomenology if the associated $W_{R}^{\pm}$, $Z_{R}$  bosons have  masses in the TeV range.
 Finally the new gauge bosons can be detected at the Large 
Hardon Collider (LHC) where ongoing experimental searches  have set the bounds $ M_{W_{R}} \geq 2.5 $ and  $M_{Z_{R}} \geq 1.162 $\cite{atlascms}. 
Over the years considerable attention has been focussed on the $SO(10)$ realisation of the left-right(LR) gauge symmetry, $SU(2)_{L}\times
SU(2)_{R} \times U(1)_{B-L}\times SU(3)_{C}$, around the TeV scale with manifest  LR symmetry  which is always accompanied by the left-right
discrete symmetry $(g_{2L}= g_{2R})$ or without it $(g_{2L}\neq g_{2R})$ and this latter symmetry is denoted as $G_{2213A}$\cite{cmp}. \\

    The purpose of this work is to show that these  gauge boson masses 
    can be realised quite effectively in the minimal symmetry breraking chain of 
$SO(10)$ or $E_{6}$ GUT:
{\small{\bf
\begin{eqnarray}
 SO(10),  E_{6}\nonumber\\ \displaystyle_{\longrightarrow}^{M_{U}} & SU(2)_{L}\times SU(2)_{R} \times U(1)_{B-L}\times SU(3)_{C} \nonumber \\
                              \ & (\equiv G_{2213A})\nonumber \\
                 \displaystyle_{\longrightarrow}^{M_{R}}  & SU(2)_{L}\times U(1)_{Y}\times SU(3)_{C}(\equiv G_{213}).
\end{eqnarray}}}
 To accunt for tiny neutrino masses we use inverse seesaw formula
 \cite{rnm-valle} for which we include three additional singlet fermions
   $S_{i}(i=1, 2, 3)$,  one  per each generation in case of $SO(10)$ but
   they are parts of standard fermion representations $\bf {27_{F_{i}}}$ of $E_6$.\\
   
    We now give some details in $SO(10)$ by using Higgs representations
$\Phi_{210}$ in the first step, ~$\chi_{16}$ in the second step, and
$H_{10}$ in the third step to achieve the low-energy symmetry.   
 In addition to conventional renormalizable interactin, we also
 include the effect of non-renormalisable ${\rm dim}.5$
 operator \cite{pp} induced by gravity effects for which $M_C \sim M_{Planck}$
 
\begin{equation}
 \mathcal{L}_{NR}=\frac{C}{M_C} Tr(F_{\mu\nu}\phi_{(210)}F^{\mu\nu}),
\end{equation} 
 leading to the GUT-scale boundary conditions on the two-loop
 estimated GUT-scale gauge couplings
 \begin{eqnarray}
 \hspace{-.8cm}
\alpha_{2L}(M_{U})(1+\epsilon_{2L})=\alpha_{2R}(M_{U})(1+\epsilon_{2R})\nonumber \\
 =\alpha_{BL}(M_{U})(1+\epsilon_{BL})=\alpha_{3C}(M_{U})(1+\epsilon_{3C})=\alpha_{G},
\end{eqnarray}
where $\alpha_{G}$ is the effective GUT fine structure constant and 
\begin{eqnarray}
  \epsilon_{2R}=-\epsilon_{2L}=-\epsilon_{3C}=\frac{1}{2}\epsilon_{BL}=\epsilon,\nonumber \\
  \epsilon= -\frac{C M_{U}}{2 M_{C}}\Bigl(\frac{3}{2\pi\alpha_{G}}\Bigr)^{\frac{1}{2}},
 \end{eqnarray}
 here $M_U=$ solution to the GUT scale that includes corrections due
 to ${\rm dim. 5}$ operator and the above boudary condition emerges by beaking the GUT symmetry through
 the  VEVs of $ <\eta(1, 1, 1)>, <\eta^{\prime}(1, 1, 15)> \subset
 210_{H}$ where the quantum numbers are under $SU(2)_L\times
 SU(2)_R\times SU(4)_C$. This implies that the $SO(10)$ and the
 Pati-Salam symmetry as well as the left-right discrete symmetry are
 broken at the GUT scale by $210$ leading to $G_{2213A}$ at lower scales. 
 
 Our solutions consistent with \  sin$^2\theta_{W}(M_{Z})=0.23116\pm
0.00013$, \ $\alpha({M_{Z})}=1/127.9$ ,and $\alpha_S(M_z)=0.1184\pm0.0007$  by including only one  light bi-doubet $h(2,2,0,1)$ and one
light right-handed doublet $\chi(1,2,-1,1)$ corresponding to
$D_{h}=D_{\chi}=1$ are shown in Table. \ref{tab:soln}. The right-handed doublet $\chi(1,2,-1,1)$ breaks the symmetry $G_{2213A}\longrightarrow G_{213}$ and 
 also generates $N-S$ mixing mass term $M$ which results in the inverse seesaw mechanism for neutrino masses.
 This  also contributes significantly towards lepton flavor
 violation. The predicted proton life time for the decay $p\to e^+\pi^0$ in our model turns out to
 be in the range $\sim 10^{37}-10^{38}$ yrs. which is beyond the
 accessible ranges of  
Super-Kamiokande ($\tau_{p}(p \rightarrow e^{+}\pi^{0}) \geq
1.4\times10^{34} yrs$) and proposed investigations at Hyper-Kamiokande
~($\tau_{p}(p \rightarrow e^{+}\pi^{0}) \geq 1.3\times10^{35} yrs$).
 
\begin{center}
\begin{table*}
\caption {Predictions for RH gauge bosons in  $SO(10)$ model with gravity induced corrections for $D_{h}=D_{\chi}=1$. } 
{\small
\hfill{}
\begin{tabular}{|l|l|c|c|c|c|c|c|c|}
\hline
$\epsilon$ & $\alpha_{G}^{-1}$ & $M_{R}$(GeV) & $M_{U}$(GeV)& $\sigma=\left(\frac{g_{2L}}{g_{2R}}\right)^{2}$ & $M_{C}$(GeV) & $C=-\frac{\kappa}{8}$ 
& $\tau_p$(yrs)\\[0.8ex]
\cline{3-8}
\hline
$0.082$ & $49.6$ & $ 2.32\times10^4$ & $2.77\times10^{16}$ & $1.274 $ & $1.12 \times 10^{18}$  & $-1.36$ & $1.04\times10^{38}$  \\
$0.083$ & $49.6$ & $ 1.9\times10^4$ & $2.75\times10^{16}$ &  $1.277 $ & $1.1 \times 10^{18}$  & $-1.37$ & $1.01\times10^{38}$\\ 
  $0.084$ & $49.7$ & $ 1.6\times10^4$ & $2.73\times10^{16}$ &  $1.280 $ & $1.09 \times 10^{18}$ & $-1.38$ & $9.84\times10^{37}$ \\ 
  $0.085$ & $49.8$ & $ 1.31\times10^4$ & $2.70\times10^{16}$ & $1.284 $ & $1.08 \times 10^{18}$  & $-1.4$ & $9.55\times10^{37}$\\ 
  $0.087$ & $49.9$ & $ 9.08\times10^3$ & $2.66\times10^{16}$ &$1.290 $ & $1.05 \times 10^{18}$  & $-1.41$& $9.02\times10^{37}$ \\ [1.3ex]
\hline
\end{tabular}}
\hfill{}
\label{tab:soln}
\end{table*}
\end{center}

As noted above,this model admits inverse seesaw formula for light neutrino masses \cite{rnm-valle}
\begin{equation}
 m_{\nu}=\frac{M_{D}}{M}\mu_{S}{\Bigl(\frac{M_{D}}{M}\Bigr)}^{T}
\end{equation}
where $M=Y_{\chi}V_{\chi}$~($Y_{\chi}$=Yukawa coupling,$V_{\chi}$ is the VEV of 
$\chi_{R}^{0}$)=${\rm  diag}(M_{1},M_{2},M_{3})>> M_{D}$ and  $\mu_S$ is small $SO(10)$-singlet-fermion mass term that violates a SM global symmetry .

 The Dirac neutrino mass matrix $M_D$ is determined by fitting the extrapolated values of all charged fermion masses at the GUT scale and running it down 
to the TeV scale following top-down approach \cite{ap,dp}. We may have to use $D_h=2$ for this purpose but to achieve near TeV scale $G_{2213A}$ symmetry
$D_h=D_{\chi}=1$ is sufficient \\

 $M_{D}(M_{R}^{0})(GeV) =$
 \begin{eqnarray}
 \hspace*{-.6cm}
{\small
 \begin{pmatrix}
0.0151 & 0.0674-0.0113i & 0.1030-0.2718i \\
0.0674+0.0113i & 0.4758  & 3.4410+0.0002i \\
0.1030+0.2718i & 3.4410-0.0002i & 83.450
\end{pmatrix}.
}\nonumber
\end{eqnarray}

The heavy neutrinos in this model are three pairs of pseudo-Dirac
 fermions which  mediate charged lepton flavor violating decays with predictions on branching ratios shown in
 Table.\ref{tab:br}. The present experimental limits on branching ratios are 
 $Br(\mu\rightarrow e\gamma) \leq 2.4\times 10^{-12}$ \cite{meg}, ~$Br(\tau\rightarrow e\gamma) \leq 1.2\times 10^{-7}$ and 
 $Br(\tau\rightarrow \mu\gamma) \leq 4.5\times 10^{-8}$ \cite{belle}. For verification of model predictions,  
 improved measurements with accuracy upto $3-4$ orders are needed. 

\begin{table}[h] 
\caption {Nonunitarity Predictions of branching ratios for lepton flavor violating decays $\mu \rightarrow e\gamma$, 
$\tau \rightarrow e\gamma$ and $\tau \rightarrow  \mu \gamma$
 as a function of singlet fermion masses} 
\centering 
{\footnotesize \bf
\begin{tabular}{lccc}
\hline\hline 
 $M(GeV)$ & $BR(\mu \rightarrow e\gamma)$ & $BR(\tau \rightarrow e\gamma)$ & $BR(\tau \rightarrow \mu \gamma)$    \\ [0.2ex] 
\hline 
 $(50,200,1711.8)$ & $1.19\times 10^{-16}$ & $4.13\times 10^{-15}$ & $5.45\times 10^{-13} $   \\ 
 $(100,100,1286)$  & $1.07\times 10^{-15}$ & $2.22\times 10^{-14}$ & $2.64\times 10^{-12} $  \\ 
 $(100,200,1702.6)$  & $1.14\times 10^{-16}$ & $4.13\times 10^{-15}$ & $5.52\times 10^{-13} $   \\ 
 $(500,500,1140)$  & $1.35\times 10^{-16}$ & $1.25\times 10^{-14}$ & $1.77\times 10^{-12} $ \\ [0.2ex]        
\hline 
\end{tabular} 
}
\label{tab:br} 
\end{table}  

The inverse seesaw formula fits the neutrino oscillation data quite
well for all the three types of light neutrino mass hierarchies.
All corresponding Higgs representations being present in $E_6$ GUT,
the same approach leads to identical results but now the three
$SO(10)$ fermion singlets are in $\bf{27}_{F_{i}}$ each of which has $10$ non-standard
fermions compared to ${16}_F+{1}_F$.\\

  We have discussed a novel method of realising low mass RH gauge
  bosons in minimal GUTs accessible to LHC using gravitational
  corrections through ${\rm dim}.5$ operator.
  The model successfully accounts for the neutrino oscillation data
through inverse seesaw mechanism. The heavy fermions in the model are
pseudo Dirac particles which are
also verifiable by their trilepton signatures at LHC. 
We have obtained similar solutions with $D_h=D_{\chi}=D_T=1$ where
$T=$ a RH triplet ~$\subset {126}~\subset SO(10)$~ but, in this case,
heavy fermions are Majorana particles with interesting experimental signatures. 
Deatils of this approach would be reported elsewhere \cite{ps}.\\

\par\noindent{\bf Acknowledgment:} M. K. P. acknowledges financial
support under DST project SB/S2/HEP-011/2013 of Govt. of India.\\

\label{}




\nocite{*}







\end{document}